\documentclass[]{spie}  
\usepackage[]{graphicx}
\usepackage{float}

\title{Performance of the HgCdTe Detector for MOSFIRE, an Imager and Multi-Object Spectrometer for Keck Observatory} 

\author{Kristin R. Kulas\supit{a}, Ian S. McLean\supit{a}, and Charles C. Steidel\supit{b}
\skiplinehalf
\supit{a}University of California, Los Angeles, CA, USA 90095-1547; \\
\supit{b}California Institute of Technology, Pasadena, CA, USA 91125
}

\authorinfo{ 
Further author information: (Send correspondence to K.R.K.)\\
K.R.K.  E-mail: kkulas@astro.ucla.edu; Telephone: 1 310 794 5582\\
I.S.M.  Email: mclean@astro.ucla.edu\\
C.C.S.  Email: ccs@astro.caltech.edu}

\begin{document} 
  \maketitle 

\begin{abstract}
MOSFIRE is a new multi-object near-infrared spectrometer for the Keck 1 telescope with a spectral resolving power of R$\sim$3500 for a 0.7$^{\prime\prime}$ slit (2.9 pixels). The detector is a substrate-removed 2K x 2K HAWAII 2-RG HgCdTe array from Teledyne Imaging Sensors with a cut-off wavelength of 2.5 $\mu$m and an operational temperature of 77K. Spectroscopy of faint objects sets the requirement for low dark current and low noise. MOSFIRE is also an infrared camera with a 6.9$^{\prime}$ field of view projected onto the detector with 0.18$^{\prime\prime}$ pixel sampling. Broad-band imaging drives the requirement for 32-channel readout and MOSFIREs fast camera optics implies the need for a very flat detector. In this paper we report the final performance of the detector selected for MOSFIRE. The array is operated using the SIDECAR ASIC chip inside the MOSFIRE dewar and v2.3 of the HxRG software. Dark current plus instrument background is measured at $<$0.008 e$^{-}$ s$^{-1}$ pixel$^{-1}$ on average. Multiple Correlated Double Sampling (MCDS) and Up-The-Ramp (UTR) sampling are both available. A read noise of $<$5e$^{-}$ rms is achieved with MCDS 16 and the lowest noise of 3e$^{-}$ rms occurs for 64 samples. Charge persistence depends on exposure level and shows a large gradient across this detector. However, the decay time constant is always $\sim$660 seconds. Linearity and stability are also discussed. 
\end{abstract}



\section{INTRODUCTION}
\label{sec:intro}  

~~~~ MOSFIRE is a near-IR spectrograph with four atmospheric bands: Y(0.97-1.12$\mu$m), J(1.15-1.35$\mu$m), H (1.46-1.81$\mu$m), or K(1.95-2.39$\mu$m).  It is located at the Cassegrain focus of the Keck-1 telescope on Mauna Kea, Hawaii.  The optical design provides both imaging and multiple-object spectroscopy over a field of view (FOV) of $6.14^\prime$ x $6.14^\prime$ with a resolving power ($\frac{\lambda}{\Delta\lambda}$) of R $\simeq$ 3,500 for a slit width of $0.7^{\prime\prime}$ (2.9 pixels along the dispersion).  The detector is a sensitive 2K x 2K H2-RG HgCdTe array with a 2.5 $\mu$m cut-off.  A unique feature of MOSFIRE is that its multiplex advantage of up to 46 slits is achieved using a cryogenic Configurable Slit Unit, or CSU, which has been developed in collaboration with the Swiss Center for Electronics and Micro Technology (CSEM).  The CSU is reconfigurable under remote control in less than 6 minutes without any thermal cycling of the instrument.  Slits are formed by moving opposing bars from both sides of the focal plane.  An individual slit has the length of $7.0^{\prime\prime}$, but bar positions can be aligned to make longer slits.  When the bars are moved apart to their full extent and the grating is changed to a mirror, MOSFIRE becomes a wide-field imager with a fine pixel scale of $0.18^{\prime\prime}$ per pixel.

MOSFIREs optical design is predicated on the 18 $\mu$m pitch of the Teledyne H2-RG detector.  High quantum efficiency (QE $\ge 65 \%$) is required to achieve our instrument throughput goals.  The on-sky measured detection rates between OH lines in the darkest parts of the Y, J, H, and K bands are, respectively, 0.3, 0.3, 0.6, 0.4 $e^{-}$/s/pixel for a $0.7^{\prime\prime}$ slit assuming $\sim$3 pixels sampling in the dispersion direction (R$\simeq$ 3,500) for optimistic conditions.  A dark current of $<$0.03 $e^{-}$/s/pixel and an effective read noise below $\sim5 e^{-}$ would result in background-limited performance for exposures longer than 90 seconds.  A typical spectroscopic integration time is 180 seconds in the K band based on the experience of the MOSFIRE team with the OH variability.  As stated in the MOSFIRE Design document, the limiting magnitudes (Vega) for a S/N ratio of 10 in 1000 seconds in spectroscopic mode at R$\simeq$3500 are J = 20.4, H=20.1, and K = 18.6.  To achieve these limits we assume 0.05 $e^{-}$/s dark current, an effective read noise of 4 $e^{-}$/pixel, and the background appropriate for spectral regions between OH lines, evaluated over a 3 pixel resolution element and assuming a 0.5 arcsec$^{2}$ extraction aperture.  Imaging limits assume a 3 x 3 pixel (0.54$^{\prime\prime}$) aperture for a point source under good seeing conditions \cite{spie2010}.

When it was selected for MOSFIRE in 2006, the H2-RG and especially the SIDECAR ASIC were recently developed technology from Teledyne Imaging Sensors. It was not known for certain at that time if the ASIC could achieve the kind of low noise performance typical of other controllers. To mitigate the risk and demonstrate that the detector and ASIC would meet the MOSFIRE sensitivity requirements, we undertook extensive testing of two engineering-grade devices and our science-grade detector. We worked closely with Teledyne on several iterations of ASIC hardware and software to achieve the final configuration.  The engineering-grade detectors were tested over a period of approximately two years.  The science-grade detector was then tested for an additional one and a half years before MOSFIRE achieved \textquotedblleft first light" at Keck.  Our results show that the MOSFIRE detector has very low dark current, low noise, excellent quantum efficiency across our wavelength range and good uniformity. It is also a very flat device, which is important for MOSFIREs fast camera beam. Reference pixels are effective in eliminating the readout shifts between the 32 outputs. The only behavior that may limit performance is charge persistence (residual images) from previous exposures. In the MOSFIRE detector there is an order of magnitude change in the charge persistence levels across the array. Our tests and analyses are described in this paper.

\section{The MOSFIRE Detector}

~~~~ The HAWAII 2-RG (\textbf{H}gCdTe \textbf{A}stronomy \textbf{W}ide \textbf{A}rea \textbf{I}nfrared \textbf{I}mager with \textbf{2}K x 2K resolution, \textbf{R}eference pixels and \textbf{G}uide mode) is designed to operate in a number of different  performance modes.  Lowest read noise is achieved with the 100 kHz \textquotedblleft slow" data rate, which is used for MOSFIRE.  The H2-RG detector can be read using 1, 4 or 32 outputs. The first uses the lowest power, but the latter provides the highest frame rates. Because MOSFIRE is also a broad-band imager, fast readout rates are essential. The delivered configuration, and the one used for almost all testing, employed 32 readout channels.  With overhead, v2.3 of the HxRG code provided by Teledyne yields a minimum exposure time of 1.45 seconds, or a frame rate of $\sim$0.7 Hz, when using 32 channels.

\begin{figure}[b]
  \begin{center}
   \begin{tabular}{cc}
   \includegraphics[height=6.5cm]{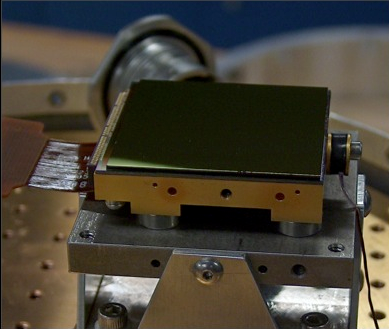}
   \includegraphics[height=6.5cm]{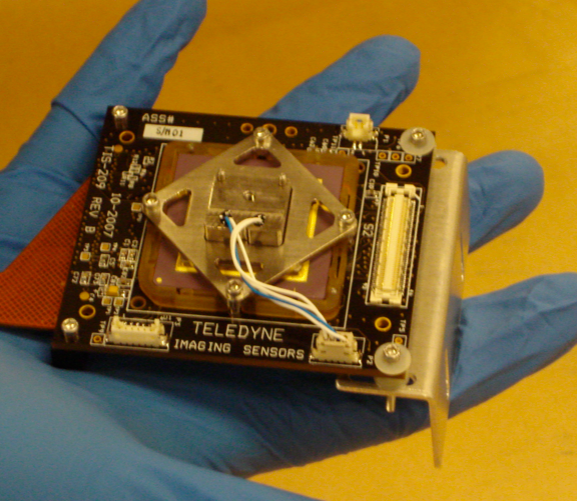}
   \end{tabular}
   \end{center}
   \caption[detector] 
   { \label{fig:detector} 
(Left) One of the engineering-grade detectors in the test dewar.  (Right) The ASIC electronics is shown next to a person's hand to demonstrate the compactness of the readout system.  This allows the ASIC to be directly next to the detector and cool along with the rest of the instrument.}
   \end{figure}

The science-grade detector installed in MOSFIRE at Keck was selected based on a consistent set of tests carried out on multiple devices by Teledyne. The selected device has a quantum efficiency of 0.88 $e^{-}$/photon (88$\%$) at 1.0 - 2.4 $\mu$m and a measured flatness of 0.00437 mm.  A median read noise of 3.45 $e^{-}$ rms was achieved using 32 Fowler pairs in Teledyne's test set up; Fowler sampling is also referred to as multiple correlated double sampling (MCDS) to indicate that the process involves taking multiple, non-destructive reads at the beginning and end of the exposure before forming the difference.  All of Teledyne's testing, however, was conducted with a standard controller and not the ASIC (\textbf{A}pplication \textbf{S}pecific \textbf{I}ntegrated \textbf{C}ircuit).  It is essential to show that the ASIC can achieve the same results.

Teledyne offers H2-RG devices with an ASIC that implements all of the detector readout functions.  The ASIC, called SIDECAR ( \textbf{S}ystem for \textbf{I}mage \textbf{D}igitization \textbf{E}nhancement \textbf{C}ontrol And \textbf{R}etrieval),  provides clocks and bias voltages to the detector and digitizes the detector outputs.  The SIDECAR ASIC is packaged separately on a small board that is located inside the dewar next to the detector head.  The ASIC provides a selection of gain settings and 16-bit digitization so that the signals leaving the dewar are already in digital form.  No external amplifiers, level shifters or Analog-to Digital Converters (ADC) are required. Weight and power are significantly reduced compared to standard controllers.  In MOSFIRE, the ASIC cryo board is at the same temperature as the bulkhead or optical bench (120 K).  Just outside the dewar wall is the JADE 2 (\textbf{J}WST \textbf{A}SIC \textbf{D}rive \textbf{E}lectronics 2) board, which provides the interface between the ASIC and USB 2.0 connection.  The stimulus for developing the ASIC/JADE 2 combination was the need to reduce power and eliminate long cables for the James Webb Space Telescope, which will also use the H2-RG devices.

\section{Lab Results}

\begin{figure}[b]
  \begin{center}
   \begin{tabular}{c}
   \includegraphics[height=8cm]{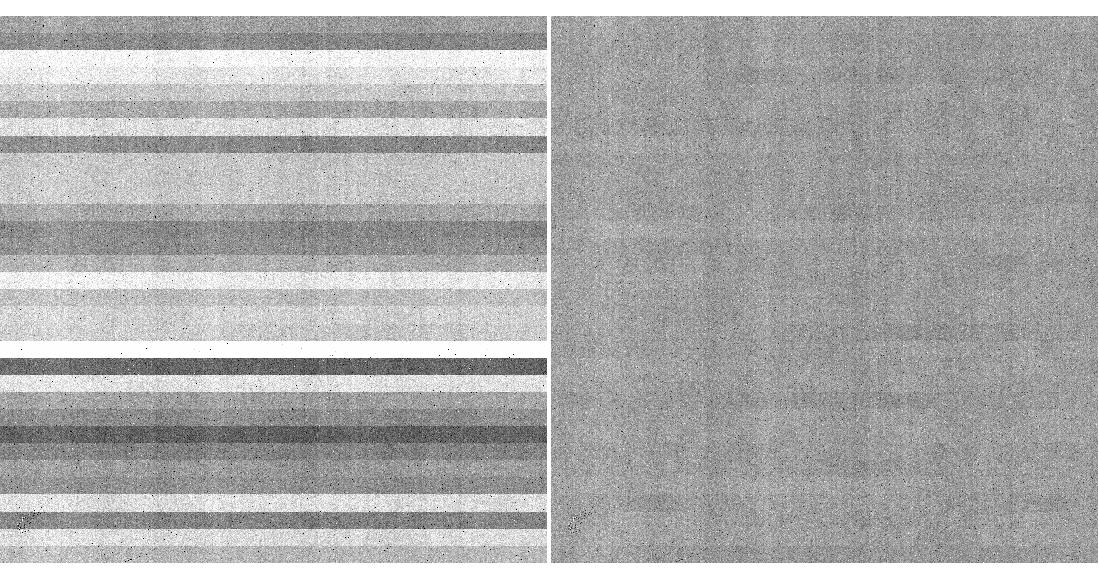}
   \end{tabular}
   \end{center}
   \caption[detector] 
   { \label{fig:refpix} 
(Left) 3 second CDS image of the science-grade detector before reference pixel subtraction.  (Right) Image after reference pixel subtraction has been implemented.}
   \end{figure}

~~~~ Numerous tests were carried out on engineering-grade detectors and our science-grade device to develop our Detector Server software, establish that the ASIC meets our requirements, and characterize the performance before MOSFIRE was installed on the Keck telescope.  Many properties were tested in several configurations to discover the best operating setup for MOSFIREs specific science goals.  There were three different lab configurations that the MOSFIRE detectors were tested in.  The first configuration consisted of a small test dewar located at the UCLA Infrared Lab, which was used for preliminary assessment of the detector, ASIC, and JADE 2.   After extensive testing was completed inside of the test dewar, the detector was then installed into the MOSFIRE detector head and the entire module, including a field-flattening lens and detector focus mechanism was cooled and evaluated inside of a larger test dewar.  The final configuration for testing was completed inside of the fully-integrated MOSFIRE instrument. 

\subsection{32-Channel Readout}

~~~~ MOSFIREs imaging capability requires the use of the 32-channel readout configuration to enable a fast readout time.  Every channel is read out from a different amplifier, each of which has a small offset bias that changes from readout to readout.  To compensate for the fluctuations in each channel, Teledyne incorporated reference pixels (the R in H2-RG) in a 4 pixel boarder around the detector.  These pixels, which are not sensitive to light, track the amplifier offsets.  The reference pixels at the top and bottom of each channel (4$\times$64 pixels each) can be used to resolve the offset issue.  We have implemented a simple algorithm that takes the median value of the top and bottom reference pixels in each channel and then subtracts that median value from each pixel in the respective channel.  Figure \ref{fig:refpix} demonstrates how channel-to-channel variation can be rectified with the use of reference pixels.

\subsection{Read Noise and Dark Current}
\label{sec:noise}  
  \begin{figure}[b]
   \begin{center}
   \begin{tabular}{c}
   \includegraphics[height=9cm]{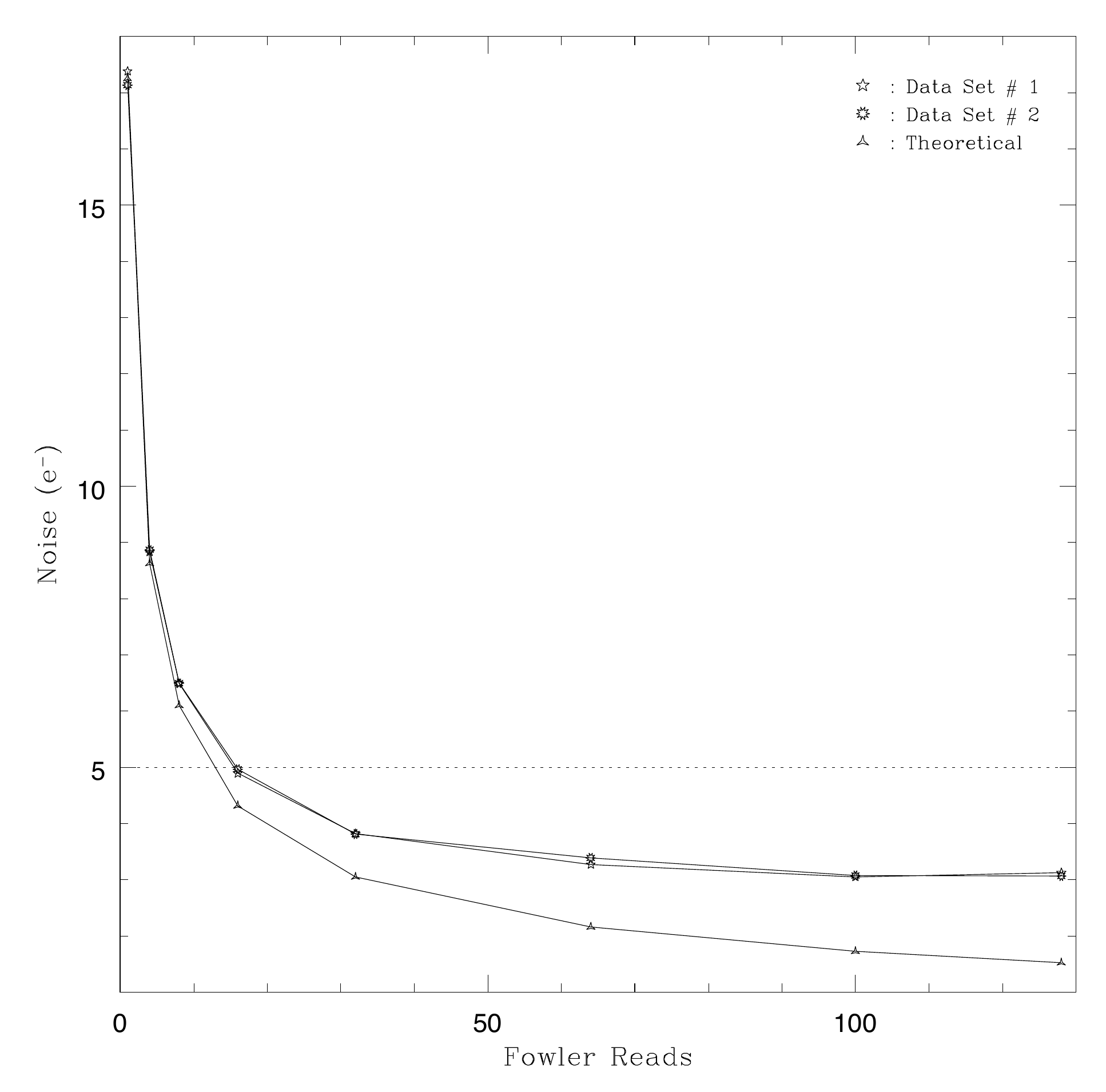}
   \end{tabular}
   \end{center}
   \caption[detector] 
   { \label{fig:readnoise} 
Read noise plot for the MOSFIRE detector using the SIDECAR ASIC readout electronics.  The read noise levels reach below 5 $e^{-}$ by MCDS 16.  The dotted line shows the 5 $e^{-}$ level.}
   \end{figure}

~~~~ Teledyne's testing of the H2-RG has shown extremely low read noise levels, below 5 $e^{-}$, can be reached.  Teledyne conducted their tests using standard controllers and not the SIDECAR ASIC.  One of our main goals was to verify that the ASIC system worked as well as a standard controller and could be used to reach the low levels of read noise obtained by Teledyne.  MCDS exposures were used for read noise tests.  Theoretically, MCDS readout will decrease the read noise by a factor of $1/\sqrt{N}$, where N is the number of reads.  For clarification,  an MCDS 16 exposure will have 16 reads taken at the beginning and 16 reads taken at the end.  This should reduce the noise by a factor of 4 from the fiducial read noise value.  Figure \ref{fig:readnoise} shows our results, with the MOSFIRE detector reaching a read noise of 4.9 $e^{-}$ with an MCDS 16 exposure.  By 64 reads the noise has reached a level of 3 $e^{-}$.  There are two different data sets shown in Figure \ref{fig:readnoise}, as well as what would be expected theoretically from the assumption that the noise would decrease as $1/\sqrt{N}$.   Table \ref{tab:rn} list the results of the read noise measurements.  There is an apparent flattening of the read noise level past MCDS 100, which we conclude is likely due to $1/f$ noise becoming the dominate noise source.  Dark current was tested by comparing a long dark frame (1800 seconds) to a short dark frame (2 seconds).  A value of $<$0.008 $e^{-}$ s$^{-1}$ pixel$^{-1}$ was measured on average.
  
\begin{table}[]
\caption{Read Noise Measurements (e$^{-}$)
\label{tab:rn}}
\begin{center}       
\begin{tabular}{|c|c|c|c|}
\hline
\rule[-1ex]{0pt}{3.5ex} MCDS Reads & Data Set 1 & Data Set 2 & Theoretical \\
\hline
\hline
\rule[-1ex]{0pt}{3.5ex} 1 & 17.4 & 17.1 & 17.3 \\
\hline
\rule[-1ex]{0pt}{3.5ex} 4 & 8.8 & 8.9 & 8.6 \\
\hline
\rule[-1ex]{0pt}{3.5ex} 8 & 6.5 & 6.5 & 6.1 \\
\hline
\rule[-1ex]{0pt}{3.5ex} 16 & 4.9 & 4.9 & 4.3 \\
\hline
\rule[-1ex]{0pt}{3.5ex} 32 & 3.8 & 3.8 & 3.1 \\
\hline
\rule[-1ex]{0pt}{3.5ex} 64 & 3.3 & 3.4 & 2.2 \\
\hline
\rule[-1ex]{0pt}{3.5ex} 100 & 3.0 & 3.1 & 1.7 \\
\hline
\rule[-1ex]{0pt}{3.5ex} 128 & 3.1 & 3.1 & 1.5 \\
\hline
\end{tabular}
\end{center}
\end{table}

\subsection{Gain Setting and Linearity}
 \begin{figure}[b]
   \begin{center}
   \begin{tabular}{c}
   \includegraphics[height=5.5cm]{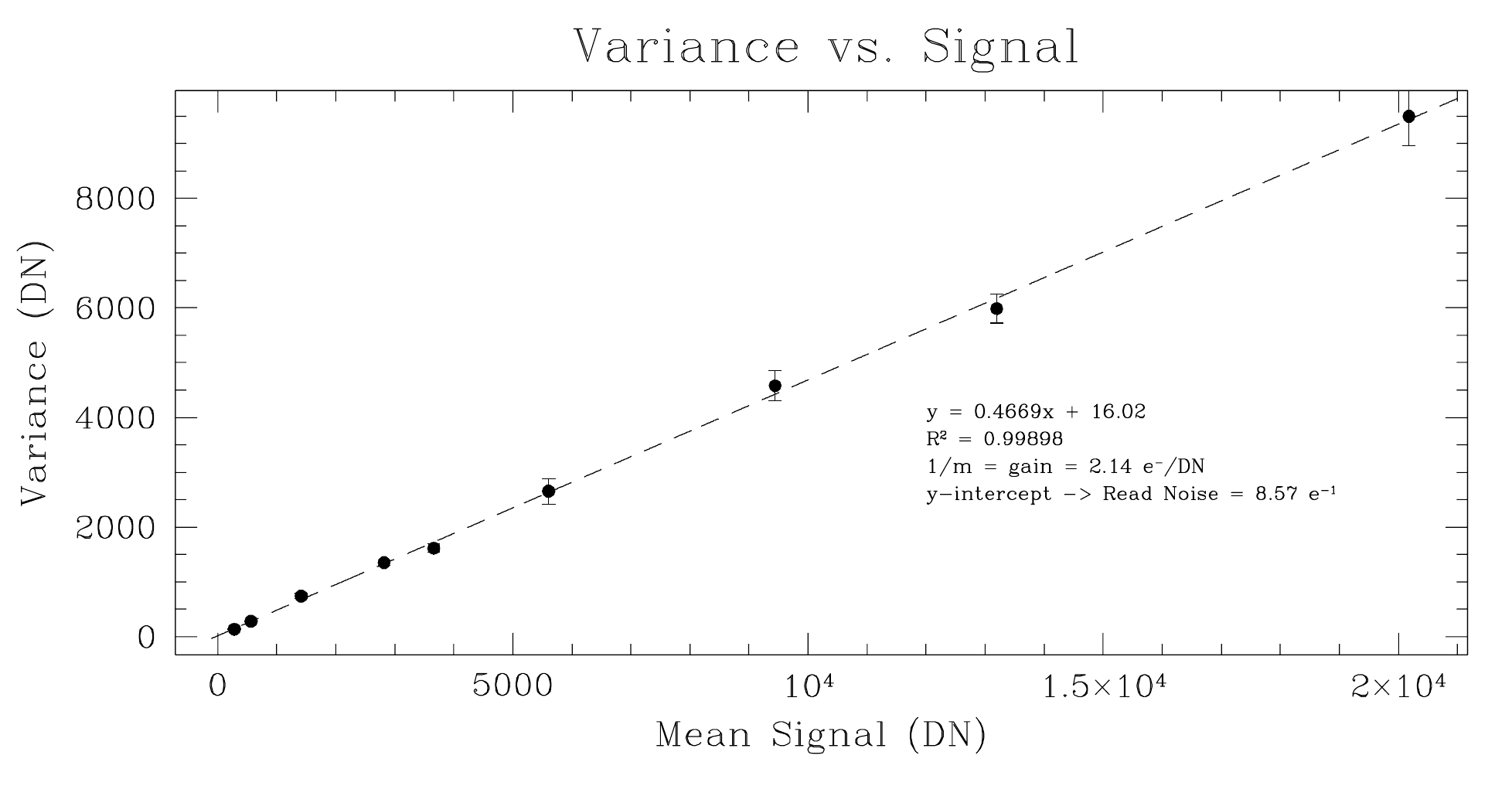}
   \end{tabular}
   \end{center}
   \caption[detector] 
   { \label{fig:gain} 
The \textquotedblleft variance method" used for estimating the detector gain.  By using a sequence of exposures with increasing signal, the gain can be calculated from the inverse of the slope of the variance versus signal.  The value calculated from this method (2.14 $e^{-}$/DN) is in very good agreement with Teledyne's value of 2.15 $e^{-}$/DN and is within less than $5\%$ with the estimated value (2.05 $e^{-}$/DN).}
   \end{figure} 

~~~~ With our current settings, Teledyne reports a total gain of 2.15 $e^{-}$/DN for the MOSFIRE detector.  To verify this number there are two different methods that can be used.  The first is to estimate the gain using the following equation~\cite{mclean2010}:

\begin{equation}
\label{eq:gain}
g=\frac{V_{fs} C}{2^{n} A_{g} e}
\end{equation}

Where $V_{fs}$ is the full-scale voltage swing allowed on the A/D unit, 3 V for the current configuration.  C is the detector capacitance at $\sim40\times10^{-15}$ farads, \textit{n} is the number of bits, which is 16, and $e$ is the charge of an electron.  $A_{g}$ is the total gain product.  MOSFIRE uses the gain setting of 15.05 dB.  To obtain the total gain product the conversion in Equation \ref{eq:db} can be used.

\begin{equation}
\label{eq:db}
\mbox{dB} = 20\: log_{10}(A_{g})
\end{equation}

The resulting value is 5.66 for $A_{g}$.  By using all of the above values with Equation \ref{eq:gain}, the estimated gain is 2.05 $e^{-}$/DN.  Additionally, the gain can be measured using an independent method known as the \textquotedblleft variance method" or \textquotedblleft photon transfer method", which uses a sequence of exposures with increasing signal.  No assumptions need to be made about the capacitance or gain, which is a benefit of this method.  By plotting the variance versus the signal from this data set, the inverse slope of the corresponding line will result in the gain of the detector.  There are two noise sources that contribute to these measurements: the photon noise of the signal photoelectrons ($p$) and the readout noise from the detector output amplifier ($R$).  The two noise sources will add in quadrature, $(noise)^{2}=p^{2}+R^{2}$.  The units of this equation are in photoelectrons, so the gain must be used to convert to DN.  This leads to the equation:

\begin{equation}
V=\frac{1}{g}S+[\frac{R}{g}]^{2}
\end{equation}

   \begin{figure}[h]
   \begin{center}
   \begin{tabular}{c}
   \includegraphics[height=5.5cm]{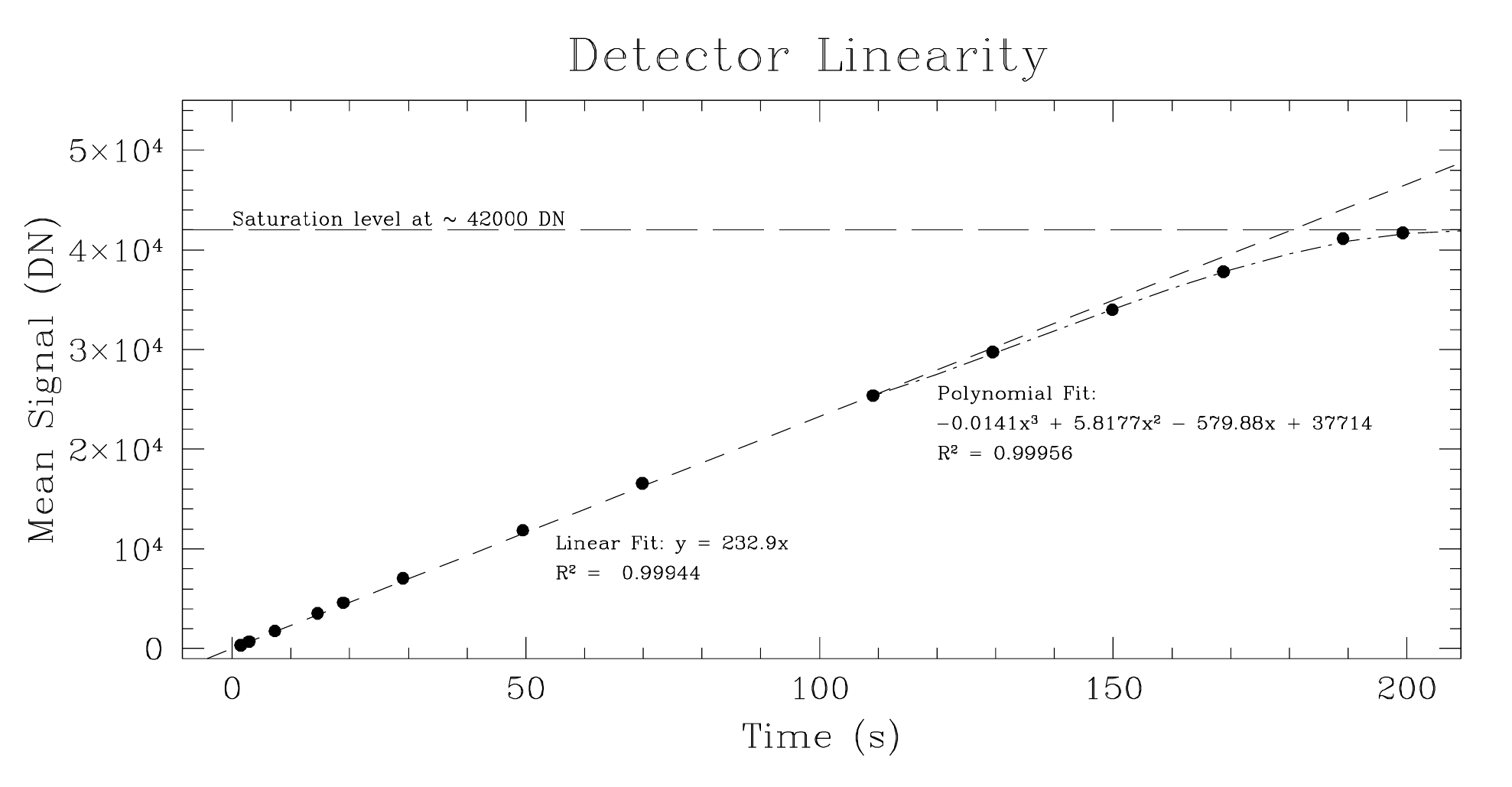}
   \end{tabular}
   \end{center}
   \caption[detector] 
   { \label{fig:linear} 
The detector was measured to hit the ADC limit at $\sim$42,000 DN, which corresponds to $\sim$90,000 e$^{-}$.  It has a linear response up to $\sim$80$\%$ full well depth.}
   \end{figure}

 \begin{figure}[H]
   \begin{center}
   \begin{tabular}{c}
   \includegraphics[height=6.25cm]{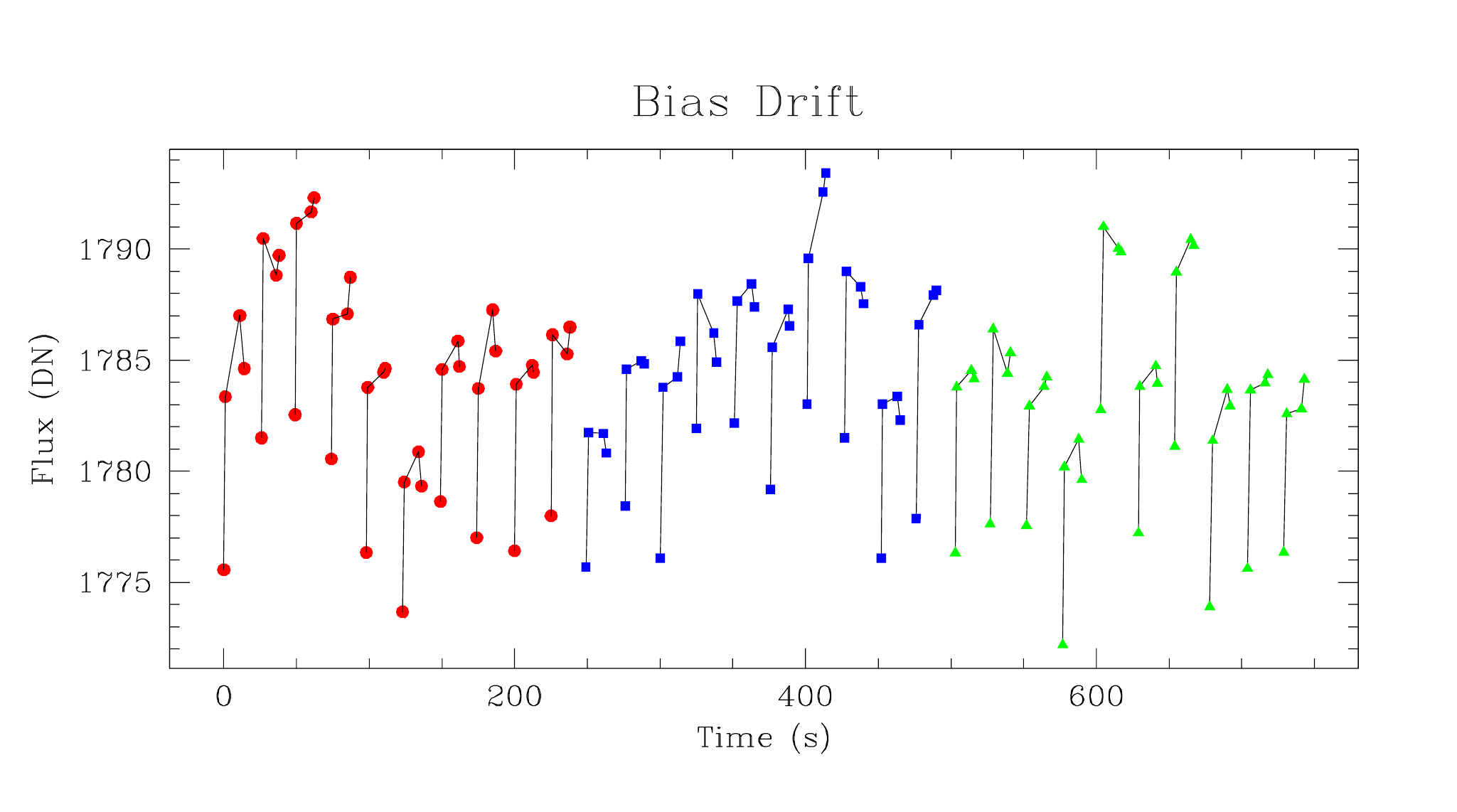}
   \end{tabular}
   \end{center}
   \caption[detector] 
   { \label{fig:bias} 
A CDS image is composed of 4 raw frames.  Here we show the 4 raw frames for a total of 30 separate CDS images.  The different colors represent three sets  of CDS images taken at different times (10 CDS images in each set).  There is an obvious drop in the readout directly after the detector is reset in all 30 frames.}
   \end{figure}

\noindent which is the equation of a straight line, where $\frac{1}{g}$ is the slope, $[\frac{R}{g}]^{2}$ is the y-intercept, and $V=(noise/g)^{2}$ is the variance.  Since $g$ can be obtained from the slope, $R$ can be solved for from the y-intercept of the line.  From this method $R$, the read noise for a CDS frame, can be independently calculated.  The measured gain is 2.14 $e^{-}$/DN (see Figure \ref{fig:gain}), which is in good agreement with Teledyne's value of 2.15 $e^{-}$/DN and is within less than a five percent difference from the estimated gain (2.05 $e^{-}$/DN).  Therefor we adopt the gain value of  2.15 $e^{-}$/DN for the MOSFIRE detector.  As well, the read noise value calculated from the y-intercept (8.57 $e^{-}$) is less than our earlier measured value of $\sim$17 $e^{-}$.  This difference can be attributed to the large error on the variance.

The linearity and saturation level of the detector was measured from the same data set used for the gain calculation.  The saturation level at the ADC limit was measured at $\sim$42,000 DN, which corresponds to $\sim$90,000 $e^{-}$. The detector was measured to be linear up to $\sim$80$\%$ of full well, after which it can be fit with a third order polynomial (see Figure \ref{fig:linear}).

\subsection{Bias Drifts}

   \begin{figure}[b]
   \begin{center}
   \begin{tabular}{c}
   \includegraphics[height=9.75cm]{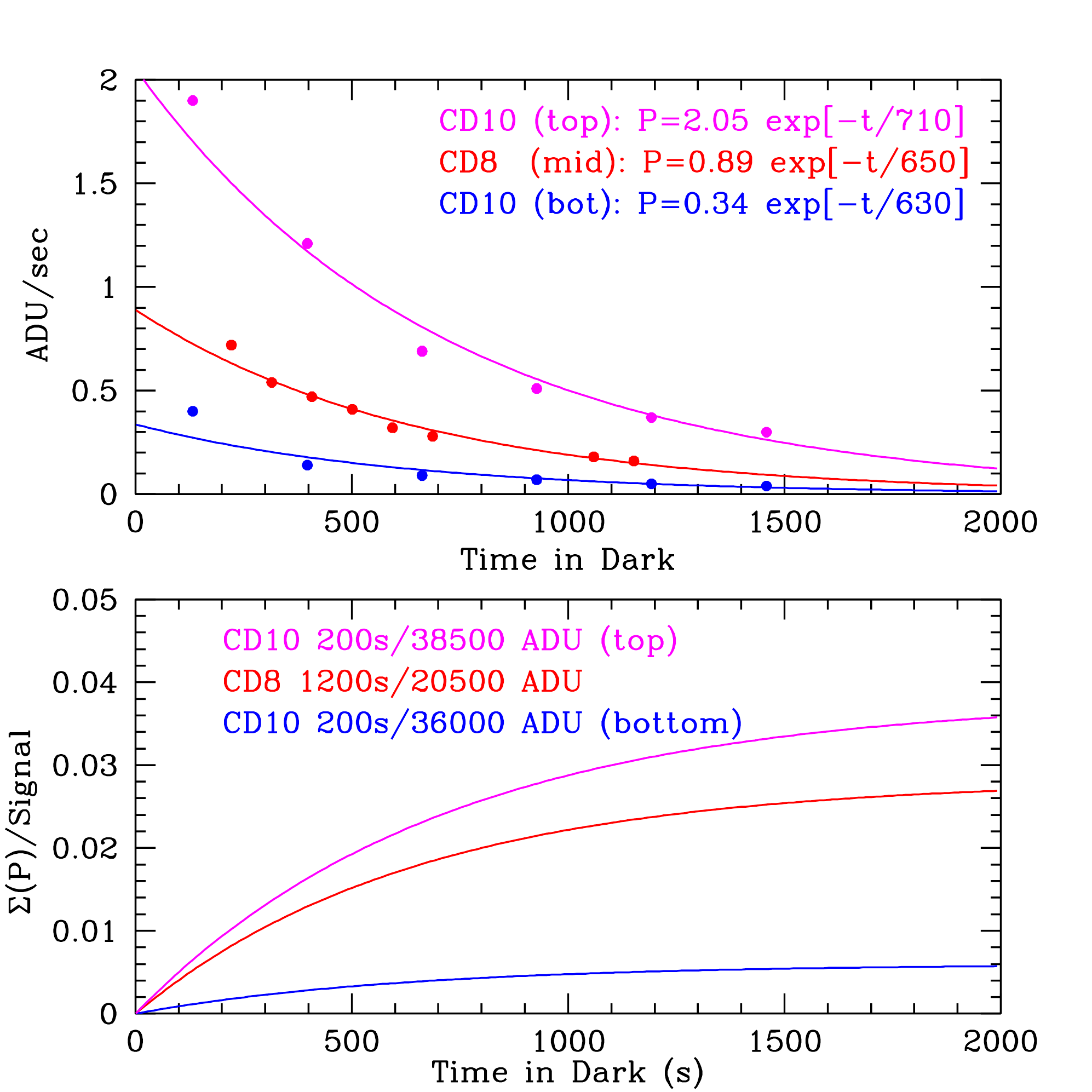}
   \end{tabular}
   \end{center}
   \caption[detector] 
   { \label{fig:decay} 
Data taken during cool down 10 and cool down 8 in the lab were used to examine persistence.  (Top)  Time after exposure versus ADU (DN) per second.  All positions in the spatial direction of the detector can be fit with an exponential function with almost the same time constant.  The amplitude of persistence, though, is much higher for the top half of the detector.   (Bottom)  Time after exposure versus  the summed persistence over the original signal.  The upper half of the detector shows much higher levels of persistence.}
   \end{figure}

~~~~In early lab testing of the science-grade detector, it was noted that CDS exposures with no illumination had negative DN counts.  A CDS image is composed of 4 raw frames.  To understand the negative DN counts, the individual raw frames were examined separately.  The results of this analysis showed a sharp drop in the readout directly after the detector is reset and that it takes approximately one readout time (1.45 seconds) before the signal settles (see Figure \ref{fig:bias}).  To resolve this anomaly an extra frame can be read and ignored at the beginning of every exposure, allowing the detector to settle to the correct level after being reset.   The added \textquotedblleft discarded" read, however, causes an unwanted overhead in the total exposure time.  The exposure type that is most affected by the bias drift is CDS.  Since MOSFIRE will primarily use MCDS mode, and the extra time overhead is not desirable, we do not use \textquotedblleft discard"  frames  following reset.  However, the option remains available in our code.

\subsection{Persistence}

~~~~Persistence is a known issue among infrared detectors\cite{smith08}.  To characterize the persistence on the MOSFIRE detector, it was illuminated to reach a value close to the ADC limit at $\sim$39,000 DN.  The detector was then put into dark and seven consecutive 200 second exposures were taken.  The residual counts measured above the inherent noise of the detector was considered to be persistence.  From this experiment, the MOSFIRE detector has a calculated decay time constant of $\sim$660 seconds, as shown in Figure \ref{fig:decay}.  With closer examination a factor of $\sim$10 difference is measured between the amplitude of the persistence along the spatial direction, perpendicular to the dispersion.  The time constant, though, appears to be consistent regardless of the level of illumination.

\section{On-Sky Results}
~~~~ MOSFIRE obtained \textquotedblleft first light" on April 4$^{th}$, 2012.  Many engineering tasks were completed to verify performance of the entire MOSFIRE system, including the detector, over four separate commissioning runs from April to June 2012.  During this time the MOSFIRE team also optimized observing procedures and observed science targets to determine practical performance limits.

 \begin{figure}[h]
   \begin{center}
   \begin{tabular}{c}
   \includegraphics[height=9cm]{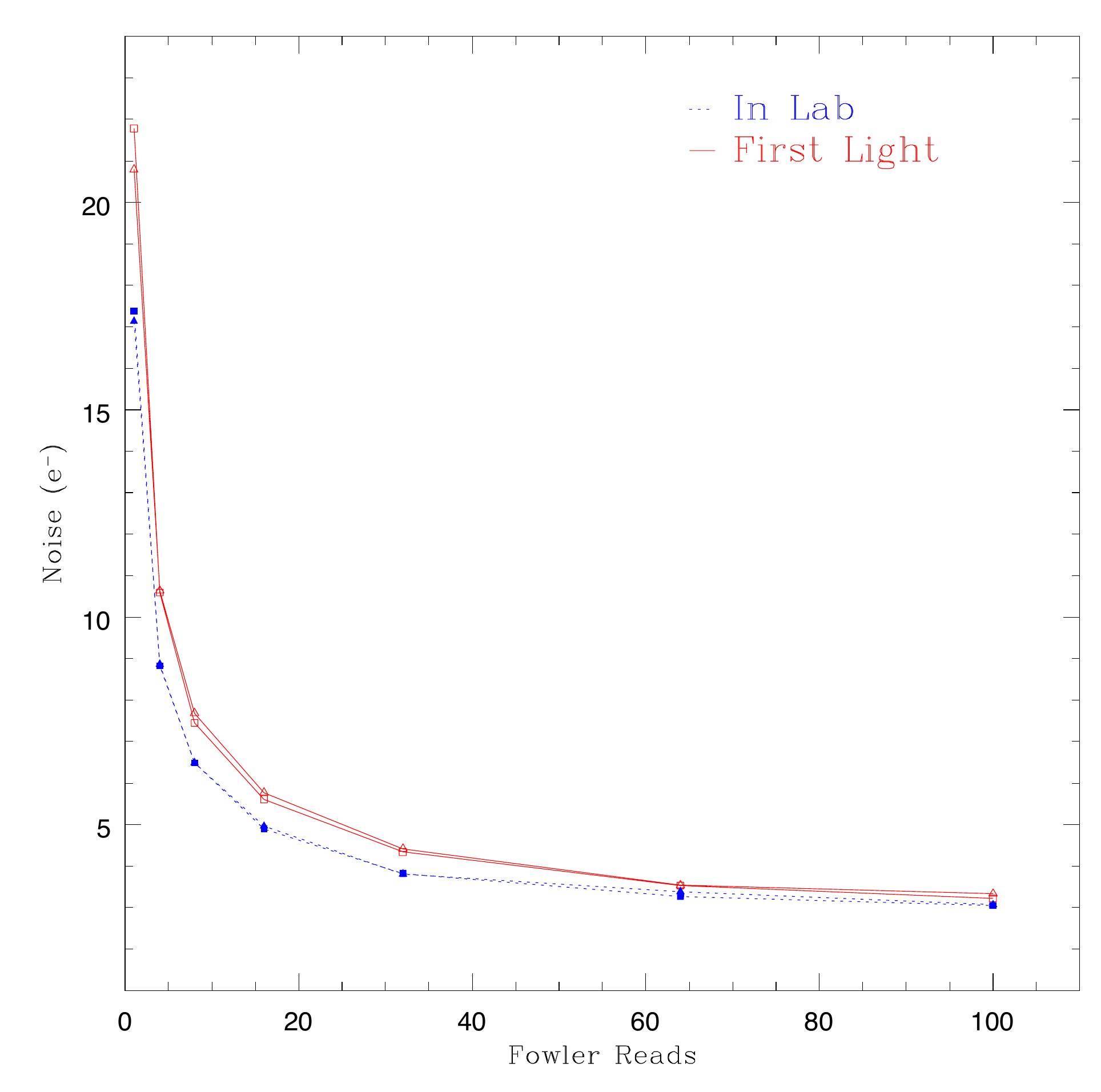}
   \end{tabular}
   \end{center}
   \caption[detector] 
   { \label{fig:noise2} 
Read noise comparison between in-lab and on the telescope measurements.  The noise is slightly worse for CDS readout mode at the telescope, but it becomes comparable to in-lab data when longer MCDS readout is used.}
   \end{figure}

\subsection{Read Noise}
~~~~ At the telescope many new noise sources can be introduced.  When dark images were taken, a noise banding was noticed on the detector that appears perpendicular to the read out channels and which was previously not seen in the lab.  Read noise tests were repeated in the same manner as discussed in Section \ref{sec:noise}  .  The read noise increased by approximately 3 $e^{-}$ for CDS mode, but noise levels became comparable to in-lab testing once MCDS mode was used (see Figure \ref{fig:noise2}).   The banding must be variable in the time domain in order for MCDS readout to average over the extra noise.  This small level of extra noise will not affect MOSFIREs performance.

\subsection{Science Verification}
~~~~ Optimizing the performance of the MOSFIRE detector was extremely important to be able to obtain the best possible scientific data.  MOSFIREs first commissioning run took place on April 4$^{th}$, 2012.  During the subsequent commissioning runs, the MOSFIRE team began science verification of the instrument.  Both galactic and extragalactic targets were observed in imaging and spectroscopic modes.

\begin{figure}
   \begin{center}
   \begin{tabular}{c}
   \includegraphics[height=12cm]{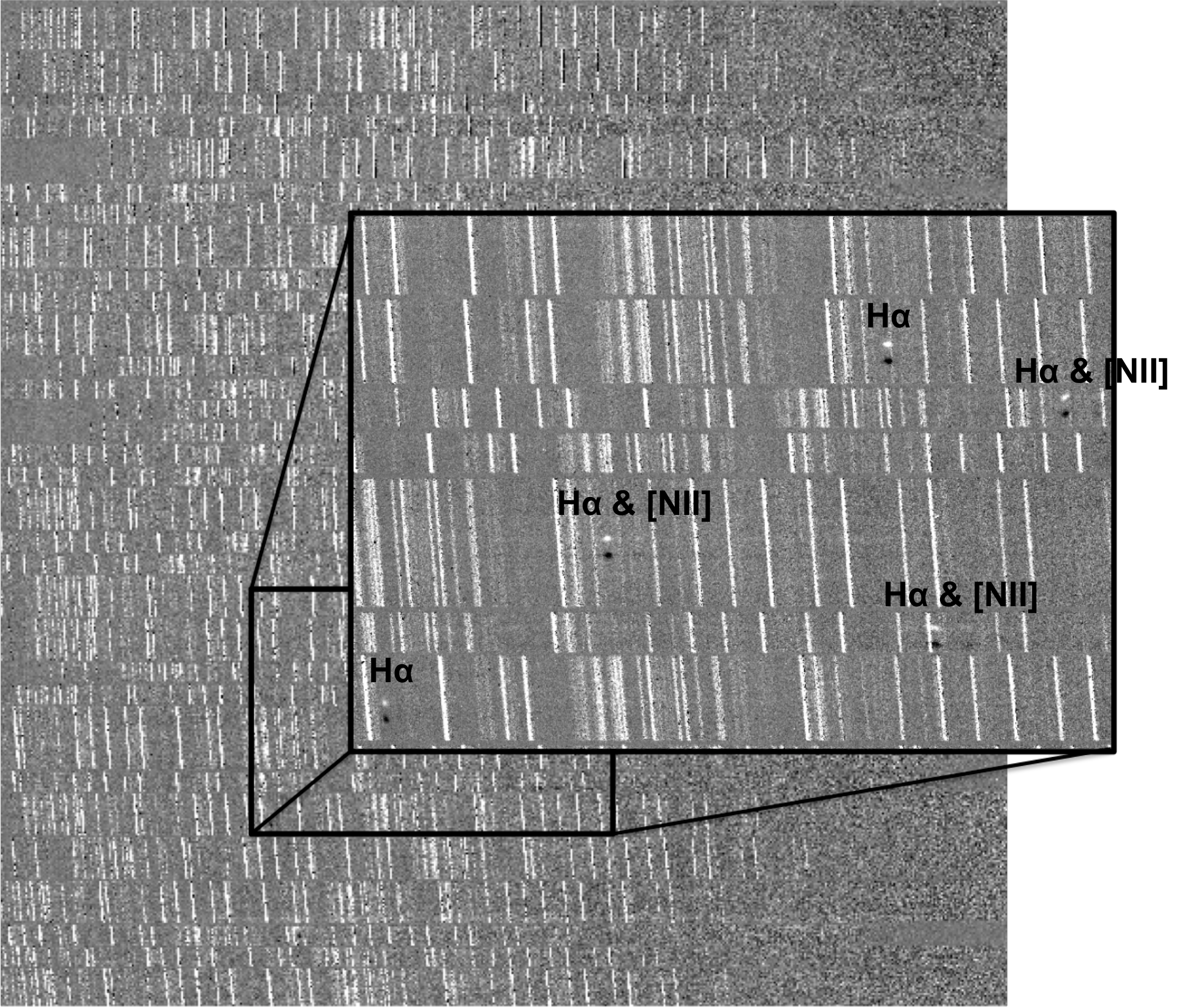}
   \end{tabular}
   \end{center}
   \caption[detector] 
   { \label{fig:spectra} 
MOSFIRE K-band spectra from a multi-object mask observing an extragalactic field.  The targets for this mask were faint ($\cal{R}$$\sim$24.4), $z\sim2$ galaxies.  The majority of the objects on the mask have H$\alpha$ detections.  A portion of the mask is magnified to show a few of the detected H$\alpha$ emission lines for individual objects, as well as several [NII] emission-line detections.}
   \end{figure}

Several masks of faint, high-redshift ($z\sim2-3$) galaxies were observed.  For a typical K-band exposure, with a measured sky background of 17.32 (Vega), a 180 second integration time was shown to be optimal for subtracting OH lines.  An ABAB nodding scheme was employed with a nod size of $\pm$1.5$^{\prime\prime}$.  Shown in Figure \ref{fig:spectra} is a stacked, differenced image of AB pairs.  A section of the masks has been magnified.  The total integration time of the stack shown is 6840 seconds (1.9 hours).  The majority of objects on the mask have H$\alpha$ detections.  Other emissions lines, such as [NII] and [SII], were were also detected. 

\section{Summary}
~~~~The H2-RG selected for MOSFIRE has proven to be an outstanding device in many respects. Measurements have shown low noise, low dark current, high QE and good uniformity. The chip is very flat and we have demonstrated that the point spread function is uniform across the field of view. Charge persistence may cause a problem, especially when changing from imaging to spectroscopic mode within a single night. Users will need to exercise care and allow sufficient decay time when changing modes. MOSFIRE will be available to the Keck community after August 2012 \cite{spie2012}.

\bibliography{report}  
\bibliographystyle{spiebib}

\end{document}